\documentclass[useAMS,usenatbib,usegraphicx]{mn2e}

\title[High resolution Br$\gamma$ spectro-interferometry of the transitional Herbig Ae/Be star HD 100546]{High resolution Br$\gamma$ spectro-interferometry of the transitional Herbig Ae/Be star HD 100546: a Keplerian gaseous disc inside the inner rim.\thanks{Based on observations made with ESO Telescopes at the La Silla Paranal Observatory under programme ID 093.C-0339}}

\author[I. Mendigut\'\i{}a et al.]
{\parbox{\textwidth}{I. Mendigut\'\i{}a$^{1}$\thanks{E-mail: \texttt{I.Mendigutia@leeds.ac.uk}}, 
W.J. de Wit$^{2}$, 
R.D. Oudmaijer$^{1}$, 
J. Fairlamb$^{1}$, 
A.D. Carciofi$^{3}$, 
J.D.~~Ilee$^{4}$ and 
R.G. Vieira$^{3}$
\vspace{0.4cm}}
\\
\parbox{\textwidth}{
$^{1}$School of Physics and Astronomy, University of Leeds, Woodhouse Lane, Leeds LS2 9JT, UK.\\
$^{2}$European Southern Observatory, Casilla 19001, Santiago 19, Chile\\
$^{3}$Instituto de Astronomia, Geof\'{\i}sica e Ci\^{e}ncias atmosf\'{e}ricas, Universidade de S\~{a}o Paulo (USP), Rua do Mat\~{a}o 1226, Cidade Universit\'{a}ria, S\~{a}o Paulo, SP - 05508-900, Brazil\\
$^{4}$SUPA, School of Physics and Astronomy, University of St Andrews, North Haugh, St Andrews, KY16 9SS, UK\\
}}

\begin{document}

\date{Accepted in MNRAS. Pre-print version}

\pagerange{\pageref{firstpage}--\pageref{lastpage}} \pubyear{0000}

\maketitle

\label{firstpage}

\begin{abstract}
We present spatially and spectrally resolved Br$\gamma$ emission around the planet-hosting, transitional Herbig Ae/Be star HD 100546. Aiming to gain insight into the physical origin of the line in possible relation to accretion processes, we carried out Br$\gamma$ spectro-interferometry using AMBER/VLTI from three different baselines achieving spatial and spectral resolutions of 2 -- 4 mas and 12000. The Br$\gamma$ visibility is larger than that of the continuum for all baselines. Differential phases reveal a shift between the photocentre of the Br$\gamma$ line --displaced $\sim$ 0.6 mas (0.06 au at 100 pc) NE from the star-- and that of the $K$-band continuum emission --displaced $\sim$ 0.3 mas NE from the star. The photocentres of the redshifted and blueshifted components of the Br$\gamma$ line are located NW and SE from the photocentre of the peak line emission, respectively. Moreover, the photocentre of the fastest velocity bins within the spectral line tends to be closer to that of the peak emission than the photocentre of the slowest velocity bins. Our results are consistent with a Br$\gamma$ emitting region inside the dust inner rim ($\la$ 0.25 au) and extending very close to the central star, with a Keplerian, disc-like structure rotating counter-clockwise, and most probably flared ($\sim$ 25$\degr$). Even though the main contribution to the Br$\gamma$ line does not come from gas magnetically channelled on to the star, accretion on to HD 100546 could be magnetospheric, implying a mass accretion rate of a few 10$^{-7}$ M$_{\odot}$ yr$^{-1}$. This value indicates that the observed gas has to be replenished on time-scales of a few months to years, perhaps by planet-induced flows from the outer to the inner disc as has been reported for similar systems. 
\end{abstract}


\section{Introduction}
\label{Sect:intro}
The Br$\gamma$ emission line at 2.166167 $\mu$m is commonly used as a quantitative accretion indicator in pre-main sequence (PMS) stars \citep{Muzerolle98,Calvet04,donehew2011,Mendi11,Mendi13}. The importance of Br$\gamma$ as an accretion tracer relies on its ubiquity in young stellar objects, and on the fact that the near-IR is much less affected by extinction than direct accretion-related UV signatures. Indeed, line modelling is able to reproduce observed Br$\gamma$ profiles mainly in terms of accretion/winds \citep{Muzerolle01,Kurosawa13,Tambovtseva14}. Due to their higher brightness and sizes when compared with classical T Tauris (CTTs), the inner discs of nearby Herbig Ae/Be (HAeBe) stars can be spatially resolved through interferometry, providing direct observational constraints on the nature of the Br$\gamma$ emission. Near-IR spectro-interferometry indicates that for most HAeBes the Br$\gamma$ emitting region is more compact than the continuum emission arising from the dusty inner disc. For some of these cases, the line has been directly associated with accretion inflows \citep{Kraus08,Eisner10,Eisner14}, but not in others where the line emitting region is more extended \citep{Malbet07,Tatulli07b,Kraus08,Weigelt11}. HAeBes have an additional interest given that they represent a class of objects where the accretion paradigm could change from magnetically channelled such as in CTTs (magnetospheric accretion; MA), to a direct disc-to star accretion through a boundary layer \citep[BL; see e.g.][]{Blondel06,Mendi11,Cauley14}.\\
HD 100546 is a nearby \citep[d $\sim$ 100 pc;][]{vandenancker98,vanLeeuwen07} transitional HAeBe (B9) star with a complex circumstellar environment. It shows an inner dust disc from $\sim$ 0.2 to 4 au, a gap from $\sim$ 4 to 13 au, and an outer disc extending several hundreds of au \citep[see e.g.][TAT11 hereafter]{Benisty10,Tatulli11}. Molecular gas traced by CO, OH and CH$^+$ accumulates mainly in the outer disc \citep{vanderplas09,Thi11,Liskowsky12,Brittain13}, but there is evidence of atomic, low density gas traced by the [O I]6300 line coming from the same spatial scale as the inner dust disc \citep{AckeAncker06}. The presence of variable emission lines in the optical/UV has been interpreted as the signature of MA/winds operating in HD 100546 \citep{Vieira99,Deleuil04,guimaraes06,Pogodin12}. The fact that there is compelling evidence that HD 100546 hosts at least two sub-stellar/planetary objects makes this system the subject of intense study \citep[see e.g.][]{Bouwman03,AckeAncker06,Quanz13,Mulders13,Brittain13,Brittain14,Walsh14}. In particular, a protoplanet is located in the outer disc, at $\sim$ 50 $au$ \citep{Quanz15,Quanz13,Currie14}, and another candidate in the disc gap at $\sim$ 10 $au$ \citep{Brittain14}.\\
Aiming to gain insight into the properties of atomic gas in the innermost regions of the disc, and their possible relations with accretion processes in HAeBe stars, we present the first high spatial resolution spectro-interferometric data of HD 100546, centred on the Br$\gamma$ line. Along this paper, the most recent stellar parameters derived from a X-Shooter/VLT spectrum \citep{Fairlamb15} will be considered:  T$_{*}$ = 9750 $\pm$ 250 K, M$_*$ = 1.9 $\pm$ 0.4 M$_{\odot}$, R$_{*}$ = 1.4 $\pm$ 0.5 R$_{\odot}$. These are consistent with previous determinations \citep{vandenancker98,guimaraes06}. Section \ref{Sect:reduction} describes the observational set-up and data reduction, section \ref{Sect:results} shows the observational results, which are analysed in section \ref{Sect:discussion} in terms of visibilities (section \ref{subsection: visibility}), differential phases (section \ref{subsect:dif_phases}) and simple modelling (section \ref{subsect:model}). Finally, section \ref{Sect:conclusions} includes a discussion, summarizing our main conclusions.  

\section{Observations and data reduction}  
\label{Sect:reduction}
HD 100546 was observed on 2014 April 16 with the VLTI/AMBER interferometer. The night was photometric, with seeing below 1$\arcsec$. Three unit telescopes (UTs) were used, covering three different configurations in terms of projected baselines (Bs) and position angles (PAs): UT4-UT1 (119--123 m, 243--258$\degr$), UT3-UT1 (79-83 m, 219--231$\degr$), and UT4-UT3 (60--61 m, 281--296$\degr$). The ranges come from the spread in the \textit{uv} coverage of the data set. The corresponding spatial resolutions, $\lambda$/2B, are between 2 and 4 mas. The high-resolution $K$-band mode with FINITO fringe tracking was used, providing a spectral resolution of $\sim$ 12000 (25 km s$^{-1}$) around the Br$\gamma$ line ($\lambda$$\lambda$ 2.15 -- 2.18 $\mu$m). The field of view is limited to the Airy disc of each individual aperture, i.e. 60 mas for the UTs in the $K$ band. Data reduction was performed using the AMDLIB-V3.0.8 software provided by the Jean-Marie Mariotti Center, following standard procedures \citep{Tatulli07a,Chelli11}. A total of 750 frames were obtained, from which we took the 20$\%$ with the best SNR to produce the final averaged observables: fluxes, visibilities and differential phases. This provides optimal results with the smallest noise, but it is noted that the final averaged data refers to frames that were taken with a maximum time delay of 1 hour, which is reflected by the ranges provided above for the B and PA values. The standard star HD 101531 (TAT11) was observed with the same configuration, before and after the observations of HD 100546, and data reduction was carried out in the same way as the science target. Instrumental artefacts were corrected for by dividing the observables of HD 100546 by those from HD 101531. This procedure also allowed us to remove telluric contamination from the observed fluxes of HD 100546. Two telluric (water) lines at both sides of Br$\gamma$ (2.163477  and 2.168687 $\mu$m) were used to perform wavelength calibration. Radial velocity correction to local standard of rest was applied. In the remainder of this paper all velocities are expressed with respect to the systemic velocity of 9 km s$^{-1}$ \citep[LSR;][]{Kharchenko07}.    

\section{Observational Results}
\label{Sect:results}
\begin{figure}
\centering
 \includegraphics[width=8.4cm,clip=true]{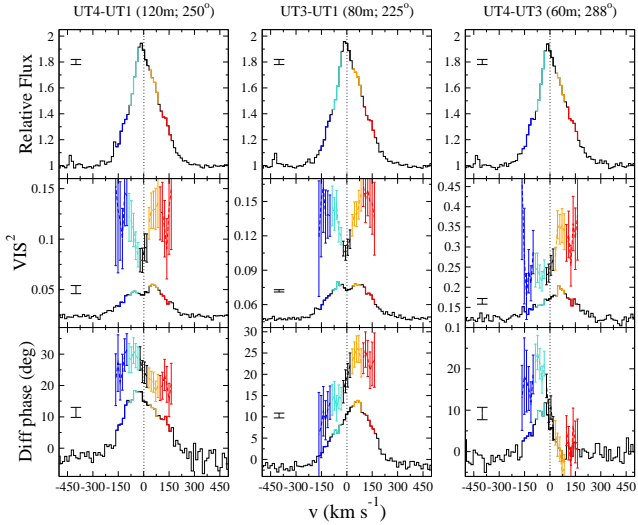}
\caption{Br$\gamma$ normalized fluxes, calibrated squared visibilities, and differential phases of HD 100546 for the baselines indicated. Vertical dotted lines show the Br$\gamma$ zero-velocity reference point for an observer comoving with HD 100546. The profiles have been coloured to indicate different spectral channels: 0 $\pm$ 20 km s$^{-1}$ (black), 60 $\pm$ 40 km s$^{-1}$ (orange), -60 $\pm$ 40 km s$^{-1}$ (light blue), 130 $\pm$ 30 km s$^{-1}$ (red) and -130 $\pm$ 30 km s$^{-1}$ (dark blue). Vertical error bars are the standard deviations of the adjacent continuum. Continuum corrected squared visibilities and phases as derived from equations \ref{Eq:Vl} and \ref{Eq:fil} (see section \ref{Sect:discussion}) are plotted above the observed ones. Their errors were derived from propagation of the individual uncertainties.}
\label{Figure:interf_results}
\end{figure}

Figure \ref{Figure:interf_results} shows the interferometric observables: Br$\gamma$ line fluxes, visibilities and differential phases. The closure phase is also measured by AMBER, showing a random noise of $\pm$ 6 $\degr$ without any clear trace of signal around Br$\gamma$. This suggests that the line emitting region is centro-symmetric on the sky, within the errors.\\ 
The total Br$\gamma$ flux (top panels of Fig. \ref{Figure:interf_results}) shows a single-peaked profile with an equivalent width (EW) of -13.6 $\pm$ 0.2 \AA{}, where the uncertainties come from the standard deviation of the measurements considering the three baselines. Because the dusty inner disc extends up to distances $\la$ 40 mas (TAT11), the field of view of our AMBER observations (60 mas) covers the spatial scale from where all the continuum in $K$ originates. Similarly, the spatial scale from which the Br$\gamma$ emission originates is fully included in the field of view, as will be argued in this paper. Therefore, the reported EW is not affected by possible lacks in the continuum or the line fluxes, representing the ``real'' EW that could be measured by any spectrograph \citep[for a wider discussion, see][]{Oudmaijer13}. The line is broadened $\sim$ $\pm$ 200 km s$^{-1}$.\\ 
Continuum visibilities were fixed to the values provided by TAT11 from $K$ band calibrated AMBER data of HD 100546. According to that work (see Fig. 1 in TAT11), the visibility of the continuum in $K$ is close to zero for our three baselines due to the (spatially extended) inner dust disc. The squared visibility increases at wavelengths corresponding to the Br$\gamma$ transition (mid-panels of Fig. \ref{Figure:interf_results}), indicating a more compact emitting region than the continuum. The baseline UT3-UT1 (and maybe UT4-UT1, considering the errors) shows a double peaked visibility with a central dip at $\sim$ 0 km s$^{-1}$. The UT4-UT3 baseline shows a single-peaked squared visibility with a redshifted maximum, although the corresponding continuum-corrected visibility (see section \ref{Sect:discussion}) is also suggestive of a double peaked profile.\\ 
The phase at Br$\gamma$ wavelengths is different than that of the adjacent continuum (bottom-panels of Fig. \ref{Figure:interf_results}), indicating a spatial displacement of the photocentre of the line emitting region with respect to that of the continuum. The differential phase peaks at different wavelengths depending on the baseline considered.  
 
\section{Analysis}
\label{Sect:discussion}
The observed line fluxes, visibilities and differential phases (F, V, $\phi$) include the contribution of the continuum (F$_c$, V$_c$, $\phi$$_c$). Following \citet[]{Weigelt11}, the visibilities and phases characterizing the pure line emitting region (V$_l$,$\phi$$_l$) are given by
\begin{equation}
\label{Eq:Vl}
F_{l}^{2}V_{l}^{2} =  F^{2}V^{2} +  F_{c}^{2}V_{c}^{2} - 2FV\cdot F_{c}V_{c}\cdot \cos \phi,
\end{equation}
\begin{equation}
\label{Eq:fil}
\sin \phi_{l} = \sin \phi \frac{ FV}{ F_{l}V_{l}},
\end{equation}
with $F_l$ = $F$ - $F_c$. The values for F$_c$ and V$_c$ are the averaged ones at both sides of Br$\gamma$. Continuum-corrected visibilities and phases are overplotted in Fig. \ref{Figure:interf_results}. In the following sections we will refer to continuum-corrected visibilities and phases. 

\subsection{Visibilities: spatial scale of the Br$\gamma$ emitting region}
\label{subsection: visibility}

\begin{figure}
\centering
 \includegraphics[width=8.4cm,clip=true]{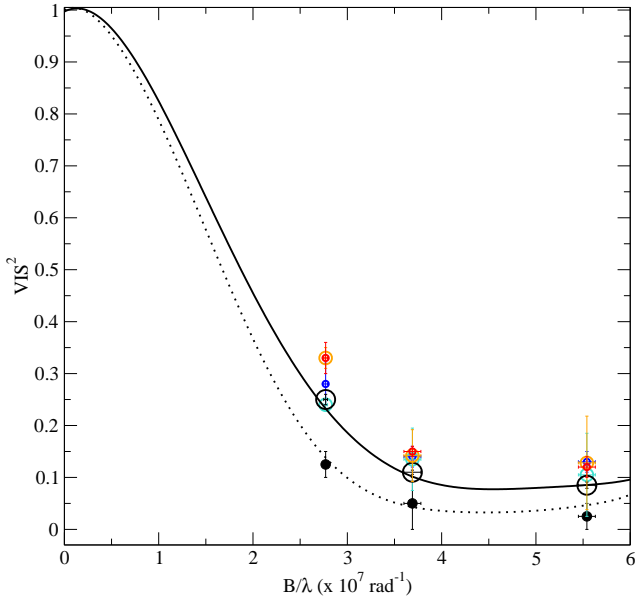}
\caption{Continuum (black solid circles), and continuum-corrected Br$\gamma$ squared visibilities (open circles; different colours representing the Br$\gamma$ spectral bins, as in Fig. \ref{Figure:interf_results}) versus the spatial frequency. The dotted line represents the fit for the continuum visibilities. Following TAT11, the brightness distribution combines a point source contributing $\sim$ 25$\%$ to the total flux, and an inclined ring contributing $\sim$ 75$\%$ ($i$ $\sim$ 33$\degr$, PA $\sim$ 140$\degr$, r$_{min}$ $\sim$ 0.25 au). The solid line represents the fit for the Br$\gamma$ visibilities of the central spectral bins. The brightness distribution combines a point source contributing $\sim$ 35$\%$ to the total flux, and an inclined disc (with the same PA and inclination, and limited by the dust inner rim) contributing 65$\%$.}
\label{Figure:visfits}
\end{figure}

Fig. \ref{Figure:visfits} shows the continuum-corrected squared visibilities against the spatial frequencies for the three baselines explored. Different velocity bins within the Br$\gamma$ emission are represented, using the same colour code as in Fig. \ref{Figure:interf_results}. Our limited \textit{uv} coverage prevents a detailed modelling of the Br$\gamma$ brightness distribution, given that different models can reasonably fit the data from only three baselines. However, we can extract some quantitative information on the relative size of the Br$\gamma$ emitting region, compared with that obtained from the $K$ band continuum visibilities (TAT11, plotted with black filled circles in Fig. \ref{Figure:visfits}). As previously mentioned, it is assumed that continuum visibilities are mainly representative of the inner dust disc. Therefore, the spatial scale of the line emitting region derived below is with respect to the inner disc, and not with respect to any other source of emission like the central star. This is based on two facts: first, the $K$ band flux primarily comes from the inner disc, as one can see in the SED shown in Fig. 2 of TAT11 (see also below and sections \ref{subsect:dif_phases} and \ref{subsect:model}). Second, and according to the same work, the visibility of the continuum in $K$ is close to 0 for our baselines, again indicating that the dominant fraction of the flux comes from a resolved source and not from the central star. The values in Fig. \ref{Figure:visfits} can be roughly fitted with simple brightness distributions (Gaussian, disc, ring, and elongated ring). In all cases the inferred Br$\gamma$ spatial scale is smaller than that of the inner dust disc, typically by a factor $\sim$ 0.8. The corresponding spatial scales associated with the red/blueshifted line emitting regions are smaller. It is noted that the line visibility level barely decreases from $\sim$ 37 to $\sim$ 55 Mrad $^{-1}$. Interpreting the line visibility as the ratio between the correlated, unresolved flux and the observed one, Fig. \ref{Figure:visfits} suggests that at $\sim$ 55 Mrad $^{-1}$ (spatial scales $\la$ 1.8 mas) there is an unresolved component contributing $\sim$ 35$\%$ to the total line flux. The Br$\gamma$ visibilities of the central spectral bins are actually more consistent with a brightness distribution that includes a point source contributing by that amount, plus a disc contributing $\sim$ 65$\%$ to the total flux (solid line in Fig. \ref{Figure:visfits}; see sections \ref{subsect:dif_phases} and \ref{subsect:model}). The disc component was assumed to have the same inclination and PA as the dust inner disc, and extends up to its inner rim ($i$ $\sim$ 33$\degr$, PA $\sim$ 140$\degr$, r$_{max}$ $\sim$ 0.25 au; see TAT11). The best model for the continuum visibilities from TAT11 is also overplotted (dotted line). \\    

\subsection{Differential phases: photocentres of the Br$\gamma$ emitting region and the continuum}
\label{subsect:dif_phases}
\begin{figure}
\centering
 \includegraphics[width=8.4cm,clip=true]{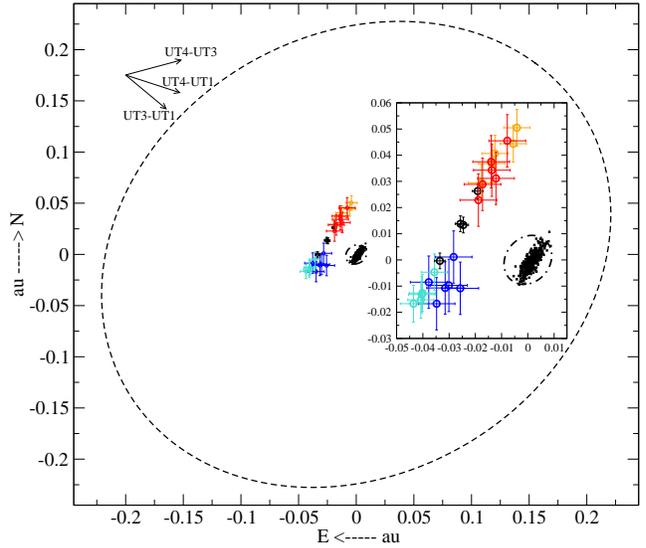}
\caption{On-sky position (angular distances were transformed to $au$ assuming a distance to the system of 100 pc) of the dusty inner rim (dashed line, from TAT11: PA = 140$\degr$, $i$ =33$\degr$, R$_d$ = 0.25 au), and an upper limit for the disc truncation radius limiting a possible MA region (dot-dashed line). The 2D-photocentre solutions (from a reference frame centred on the star) of the continuum are the black symbols inside the disc truncation radius (representing 360 velocity bins homogeneously distributed at both sides of the Br$\gamma$ line), and those of the Br$\gamma$ emission are the circles (same colour code as in Fig. \ref{Figure:interf_results}). The top left vectors show the directions of the baselines. The embedded panel is a zoom-in of the innermost regions.}
\label{Figure:indisk}
\end{figure}

Differential phases provide information of photocentre displacements along the direction of the baselines, on angular scales that can surpass the nominal resolution of the interferometer. Following \citet{Lebouquin09}, wavelength-dependent photocentre displacements can be derived from the continuum-corrected differential phases ($\phi_i$) by
\begin{equation}
\label{Eq:phot_dis}
\vec{p_i} = \frac{-\phi_i}{2\pi} \cdot \frac{\lambda}{\vec{B_i}},
\end{equation}
where $\vec{p_i}$ is the projection on the baseline $\vec{B_i}$ of the 2D-photocentre vector. The 2D-photocentre vectors at different wavelengths can therefore be recovered by solving the above system of equations. As a first order approach (see next section), it is assumed that the dust continuum emission is homogeneously distributed along the inner disc. Therefore, the symmetry of the system forces its photocentre to be centred on the (unresolved) star, which is initially assumed to be the reference frame from which photocentre displacements are measured. Fig. \ref{Figure:indisk} shows the on-sky positions of the derived 2D-photocentre displacements with respect to the central star, considering spectral bins that represent the continuum and the different components of the Br$\gamma$ line. The dusty inner rim is represented by the dashed line (TAT11). The dot-dashed line is an upper limit for the disc truncation radius (R$_{t}$ $\sim$ 0.01 $au$), i.e. the maximum stelleocentric distance from which gas can be accreted magnetospherically. This has been derived from the maximum magnetic field provided in \citet{Hubrig09}, and the expression for R$_{t}$ in \citet{Tambovtseva14} (see section \ref{Sect:conclusions}).

The 2D photocentre solutions of the continuum bins are oriented in the SE-NW direction, which is the same as the major axis of the dusty disc reported by TAT11 (PA =140 $\pm$ 16 $\degr$). A linear fit to our data confirms this result, providing PA = 139 $\pm$ 6$\degr$. In addition, the 2D photocentres of the different spectral bins of the line are also aligned (PA = 152 $\pm$ 8$\degr$) with the major axis of the system. Given that our \textit{uv} coverage is not homogeneous, it cannot be ruled out that those alignments could result from the similar directions of the three baselines covered here. Other baselines should be explored to confirm those results, preferably oriented in directions perpendicular to the ones analysed. In addition, the 2D photocentre solutions of the redshifted parts of the line are located towards the NW direction, and the blueshifted parts towards the SE. This cannot result from the previously mentioned configuration of the baselines, and constitutes an indication of a rotating, disc-like structure. Interestingly, the sense of rotation is the same as previously reported from other gas tracers in outer parts of the disc of HD 100546 \citep[i.e. counter-clockwise; see][]{AckeAncker06}. Moreover, there is a rough tendency for the high-velocity bins (red and blue circles) to be closer to the 2D photocentre of the peak emission (black circles), and for the low-velocity bins (orange and light blue circles) to be further away. This is suggestive of Keplerian rotation, although the error bars are large to provide a firm conclusion based on the observations alone (see the next section). Finally, the 2D photocentre of the Br$\gamma$ peak emission is displaced $\sim$ 0.03 au NE from the continuum photocentre, which in this section has been assumed to be in the central star.\\

\subsection{Inner dust and gas modelling}
\label{subsect:model} 
In order to further explore the photocentric shift between gas and dust emission we performed simple modelling using a Keplerian rotating gas plus a dusty inner disc. As suggested in section \ref{subsection: visibility}, the Br$\gamma$ flux consists of two components, for which we assume a rotating disc and an unresolved Gaussian emission region. The rotating component is limited by the dust inner disc at 0.25 au (TAT11), and an assumed inner hole at R$_{t}$. Although a disc wind may play a role \citep[e.g.][]{GarciaLopez15}, this is not considered in our model. The wind contribution would be resolved out and therefore it would not contribute to the visibilities and differential phases available so far. A continuum image of the inner dust rim was created using the radiative transfer code in \citet{Whitney03}, which has been previously used to model high angular resolution data of young star discs \citep[see e.g.][]{Akeson05}. The dust structure is a standard, flared accretion disc with $H(r) = H_{0}(r/r_{0})^{\beta}$ ($\beta=1.25)$, and the inner rim is assumed to be vertical (as opposed to curved). The inclination and PA for both the dusty inner disc and the Keplerian, Br$\gamma$ emitting region are the ones derived by TAT11.\\
\begin{figure}
\centering
 \includegraphics[width=8.4cm,clip=true]{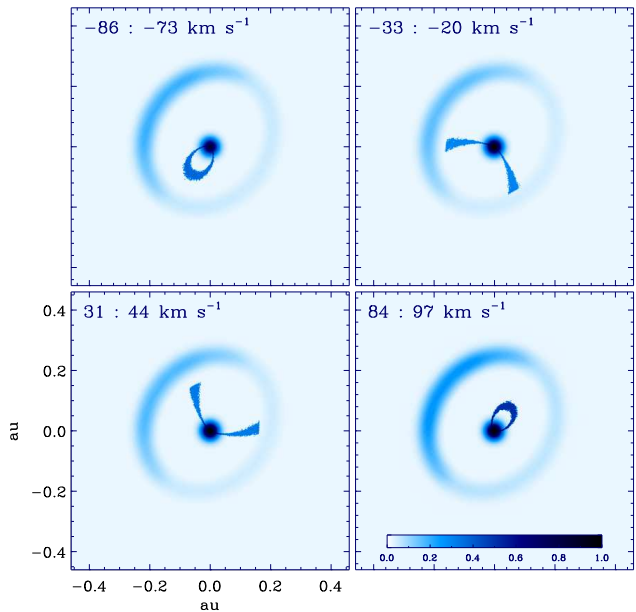}
\caption{Modelled inner dust rim and velocity surfaces of the gaseous disc around HD 100546, as described in the text. North is on the top and east is on the left. The NE part is further from us and the SW is closer. Symmetric distributions of gas at velocity ranges -86 -- -73, -33 -- -20, 31 -- 44, and 84 -- 97 km s$^{-1}$ are represented in the top left, top right, bottom left and bottom right panels, respectively. The colour gradient indicates the relative brightnesses in each panel, with the darkest colour representing the maximum flux.}
\label{Figure:modelling}
\end{figure} 
Fig. \ref{Figure:modelling} shows the result of this modelling considering different velocity bins for the gas. From the model and the inclination of the source, the inner rim in the NE region --further away from us than the SW-- dominates the total $K$ band continuum emission (see also e.g. \citet{AckeAncker06}, TAT11, \citet{Avenhaus14}, \citet{Fedele15}), which results in a continuum photocentric shift of $\sim$ 0.03 au from the central star in that direction. The photocentre displacement in each spectral channel of the Br$\gamma$ emission was calculated and converted to phase shifts by projecting on to the three VLTI baselines used. The modelled photocentre displacements and flux were convolved with a Gaussian point spread function for a spectral resolution of 12000. The results from the model are compared to the observations in Fig. \ref{Figure:model_obs}. The modelled photocentre displacements (bottom panels) are in good agreement with those observed. The peaks in the photocentre displacements correspond to the peaks of the flux component related with the disc's velocity profile, which also coincide with the two peaks shown in the visibilities of Fig. \ref{Figure:interf_results}. The modelled, correlated (unresolved) flux is smaller than the observed one, which is consistent with the low observed visibilities. Clearly, a significant fraction of the total line flux is resolved out at the baseline lengths employed. The modelling also reveals that large inner rim heights cause stronger photocentric displacements, whereas other parameters such as the inclination or the inner radius for the gas have a smaller effect. A minimal scaleheight ($H/r\sim0.02$) at the inner rim was found to best fit the differential phases, consistent with the idea that the dusty inner rim is not puffed up (TAT11).\\
\begin{figure}
\centering
 \includegraphics[width=8.4cm,clip=true]{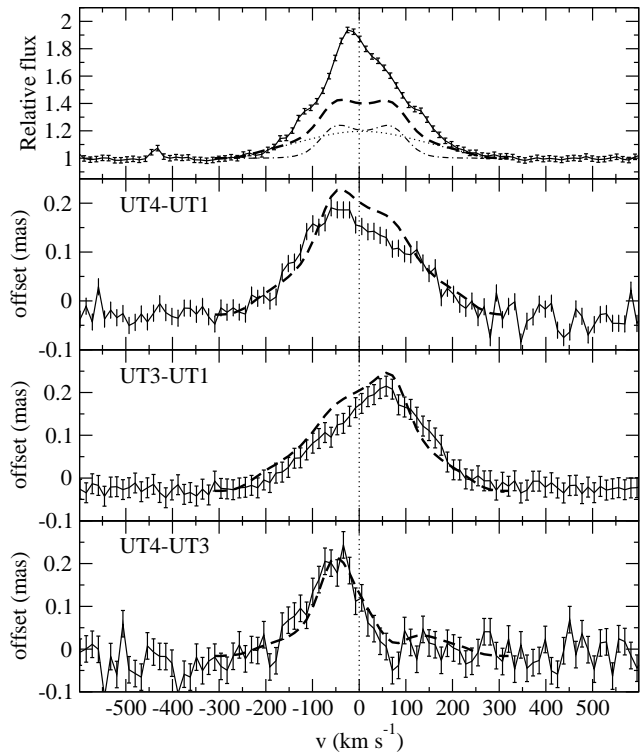}
\caption{Observed Br$\gamma$ flux and modulus of the photocentric displacements along the three baselines (solid lines with error bars) versus model results (dashed lines) as described in the text. The modelled flux consists of a Gaussian plus a Keplerian profile (dotted and dash-dotted lines in the top panel).}
\label{Figure:model_obs}
\end{figure}
In summary, the continuum photocentre of the star-disc system does not coincide with the central star --as it was assumed in the previous section-- but is displaced $\sim$ 0.03 $au$ to the NE. In addition, although the observed Br$\gamma$ profile is single peaked, the correlated flux is double peaked and the phases indicate that the resolved Br$\gamma$ emission is in Keplerian rotation and interior to the dusty inner rim.

The fact that the continuum photocentre is displaced $\sim$ 0.03 $au$ NE from the star results from the inner disc orientation, with the furthest part directly illuminated by the star. Since the continuum photocentre is the reference frame from which the Br$\gamma$ one is measured, and given that this is located an additional $\sim$ 0.03 $au$ NE from it (previous section), the photocentre of the gaseous disc traced by Br$\gamma$ is shifted by $\sim$ 0.06 $au$ from the central star in the same direction. The most plausible explanation for this shift is that the gaseous disc is flared, implying that the (furthest) NE part of it offers a larger projected surface than the SW. This interpretation is analogous to the one given by \citet{AckeAncker06} in order to explain the photocentric shift to the NE also observed in the [OI]6300 line. Despite this type of modelling is out of the scope of this work, a simple estimate of the flaring angle can be derived from geometrical considerations. A photocentre displacement of $\sim$ 0.06 $au$ to the NE requires the projected area of the NE disc to be $\sim$ 2 times the one of the SW. For a constant flaring angle $\alpha$ (defined as the angle subtended by an inclined disc with respect to a flat disc), the ratio of the further-to-nearby projected disc surface equals cos ($i$ - $\alpha$) / cos ($i$ + $\alpha$). For $i$ $\sim$ 33$\degr$, this requires $\alpha$ to be $\sim$ 25$\degr$, or $H/r$ = tan($\alpha$) $\sim$ 0.50. 

Alternatively, it could be argued that the SW part of the gaseous disc is partially obscured by the SW inner rim. However, this would need the inner rim to be significantly puffed-up, for which there is no evidence. Finally, an exciting possibility that could eventually explain the Br$\gamma$ photocentric shift is that the Keplerian gaseous disc is not centred on the star but on a different source displaced $\sim$ 0.06 $au$ NE from it. This disc could be associated with an accreting, unresolved sub-stellar/planetary companion. The corresponding rotational period around the star would be around 10 days, for which the probability that we have detected it precisely when that is located in the same direction than the bright inner rim would only be around 10-15 $\%$. Although this possibility cannot be completely discarded with the data in hand, we consider the flared Br$\gamma$ disc centred on the star as the most likely scenario. Additional baselines and uv-coverage are necessary to carry out a more detailed modelling that allows us to confirm/discard this view.

\section{Discussion and conclusions}
\label{Sect:conclusions}
We have presented the first high-resolution Br$\gamma$ spectro-interferometric data of HD 100546. Our results are consistent with a Br$\gamma$ emitting region inside the inner rim ($<$ 0.25 au) and extending very close to the star. Most of the Br$\gamma$ emission can be associated with a Keplerian, disc-like structure coplanar with the dusty inner disc, and rotating counter-clockwise (SE towards us and NW moving away from us). Simple modelling is also consistent with this view. The Br$\gamma$ photocentric shift of $\sim$ 0.06 $au$ NE from the star is most probably due to the flared geometry ($\sim$ 25$\degr$) of the gaseous disc. Additional baselines and uv coverage is needed to fully understand the nature of the Br$\gamma$ disc.\\
\subsection{Accretion in HD 100546}
\label{Sect:accretion}
The Br$\gamma$ line is a common accretion tracer in PMS stars. The (non-photospheric) Br$\gamma$ luminosity of HD 100546 can be derived subtracting a photospheric EW of $\sim$ 24 $\AA{}$ \citep[from][]{Kurucz93} to the observed EW, and then assuming that the disc continuum flux in K represents 75$\%$ of the total observed flux (TAT11). This is in turn given by a 2MASS typical K magnitude = 5.42. The Br$\gamma$ luminosity can then be associated with a mass accretion rate from the accretion-line luminosity relation in \citet{Mendi11} \citep[also from][]{donehew2011}, which was derived assuming MA operating in HAeBe stars. The mass accretion rate of HD 100546 obtained this way is $\sim$ 3.6 $\times$ 10$^{-7}$ M$_{\odot}$ yr$^{-1}$. The same result within uncertainties is obtained from MA modelling of the UV-Balmer excess by \citet{Fairlamb15}: 2.5 $\times$ 10$^{-7}$ M$_{\odot}$ yr$^{-1}$. When the UV excess is instead modelled from BL accretion, a higher value up to  2 $\times$ 10$^{-6}$ M$_{\odot}$ yr$^{-1}$ is obtained \citep{Blondel06}. However, \citet{Muzerolle04} showed that accretion rates larger than 10$^{-6}$ M$_{\odot}$ yr$^{-1}$ would make the inner gas optically thick. As discussed in that work, that would have a double effect: first, the gas would significantly contribute to the total emergent flux in the $K$ band; secondly, the stellar radiation would not heat the inner dust rim, which would therefore be located closer to the star than the dust sublimation radius. In contrast, the $K$ band flux in HD 100546 is dominated by the inner dust disc, and the location of the inner rim is consistent with the dust sublimation radius (TAT11). Therefore, the mass accretion rate estimated assuming BL \citep{Blondel06} is too high to explain the observed inner disc properties of HD 100546. In addition, the inner rim is not puffed-up according to TAT11 and our modelling. Consequently, the amount of inner gas we have detected should be large enough to block the stellar radiation and prevent the inner rim to be heated and puffed-up, but at the same time small enough to prevent the presence of dust at distances closer than the dust sublimation radius.\\

There are additional lines of evidence suggesting that accretion in HD 100546 could be magnetospheric, and not through a BL. Equation 3 in \citet{Krull99} provides a lower limit to the magnetic field necessary to drive MA, in terms of the stellar parameters and accretion rate. Assuming a mass accretion rate of $\sim$ 10$^{-7}$ M$_{\odot}$ yr$^{-1}$, the minimum stellar mass and maximum stellar radius allowed by the uncertainties provided by \citet{Fairlamb15}, and a minimum rotation period of 0.26 days \citep[from a maximum projected rotational velocity $v sin i$ = 65 km s$^{-1}$ and a minimum inclination i = 22$\degr$, from][and TAT11] {guimaraes06}, HD 100546 could require a magnetic field of only several tens of Gauss to drive accretion magnetospherically\footnote{Note that large uncertainties are involved in the calculation of the magnetic field necessary to drive MA. For the typical error bars of the stellar parameters of HD 100546, that ranges between a few tens of Gauss and a few kilo-Gauss}. This is consistent with the magnetic field measured by \citet{Hubrig09}, 89 $\pm$ 26 G, although non detections have also been reported \citep{Donat97,Hubrig13}. In addition, if accretion in HD 100546 is actually magnetospheric the Keplerian gaseous disc should be truncated at a stelleocentric radius that increases with the stellar magnetic field and radius, and decreases with the accretion rate and stellar mass \citep{Elsner77}. Using equation (6) in \citet{Tambovtseva14}, the disc truncation radius of HD 100546 should be $\la$ 0.01 au (Fig. \ref{Figure:indisk}). This value is consistent with the Keplerian radius inferred from the width of the Br$\gamma$ line profile ($\sim$ $\pm$ 200 km s$^{-1}$; section \ref{Sect:results}); once this value is de-projected using the inclination in TAT11, the corresponding Keplerian distance is $\la$ 0.02 au. It is noted that, in contrast to optically thick lines like H$\alpha$, Br$\gamma$ is mainly broadened by the Doppler effect \citep{Tambovtseva14}. If the Keplerian disc is truncated by the stellar magnetic field the gas would then fall ballistically on to the stellar surface. This could eventually be traced from the presence of redshifted absorptions at free-fall velocities in the profiles of several lines. In fact, previous spectroscopic analysis involving optical/UV lines suggest MA/ejection processes in HD 100546 \citep{Vieira99,Deleuil04}, with redshifted absorptions at velocities comparable to free-fall \citep{guimaraes06}. However, those signatures are not observed in the Br$\gamma$ profile of HD 100546. The unresolved component contributing $\sim$ 35$\%$ to the visibility (section \ref{subsection: visibility}) could be related to magnetospheric infall, but the small spatial scales involved cannot be probed from our observations.\\

\subsection{Origin of the inner disc gas}
\label{Sect:origin}
With a mass accretion rate of a few 10$^{-7}$ M$_{\odot}$ yr$^{-1}$, and a gaseous inner disc mass of $\sim$ 10$^{-8}$ M$_{\odot}$ (from the dust inner disc mass in TAT11 and a gas to dust ratio of 100), the inner disc would be depleted of gas in less than a year. Lower values for the gas to dust ratio \citep[e.g.][]{Brittain09,Benisty10} or differences within 1 dex for the accretion rate would not significantly modify that result, which indicates that the survival time of the inner disc gas can only be a few months/years. This would require that the observed gas in the inner disc is being replenished, perhaps by flows from the outer to the inner disc. This type of planet-boosted flows has been observed in similar HAeBes like HD 142527 \citep{Casassus13}, which is accreting at a similar rate than HD 100546 \citep{Mendi14}. If planets are indeed being formed in this type of systems, they do not seem to have a significant effect on the stellar accretion rates, given that 10$^{-7}$ M$_{\odot}$ yr$^{-1}$ is a typical value for HAeBe stars \citep{Mendi11}. In other words, despite planets in formation can induce gas transfers between different parts of the disc \citep{DodsonSalyk11}, the amount of gas trapped by them would be low enough to keep the stellar accretion rate practically unaltered, as expected from the comparatively low planetary accretion rates \citep{Close14,Zhou14}. It is also noted that the fact that the accretion rate associated to the Br$\gamma$ emission is several orders of magnitude larger than for sub-stellar/planetary objects argues against the unlikely explanation that the $\sim$ 0.06 $au$ Br$\gamma$ photocentric displacement could be associated to an accreting disc centred on such a close companion (section \ref{subsect:model}).\\ 
\subsection{Final remarks}
\label{Sect:final}
As argued before, and regardless from a possible MA contribution to the Br$\gamma$ line, most of its emission comes from a more extended, Keplerian rotating disc. \citet{Kraus08} found a correlation between the H$\alpha$ profile and the extent of the Br$\gamma$ emitting region with respect to that of the continuum characterizing the inner dust rim, which in turn would be related with the physical origin of the line. Stars with a P-Cygni H$\alpha$ profile show particularly compact Br$\gamma$-emitting regions (R$_{Br\gamma}$/R$_{cont}$ $<$ 0.2) more consistent with accretion, while stars with a double-peaked or single-peaked H$\alpha$ profile show a significantly more extended Br$\gamma$-emitting region (0.6 $\la$ R$_{Br\gamma}$/R$_{cont}$ $\la$ 1.4). The H$\alpha$ profile of HD 100546 is variable, but always double-peaked \citep{Vieira99} or single-peaked \citep{Fairlamb15}. This, along with the relative size of the Br$\gamma$ emitting region (R$_{Br\gamma}$/R$_{cont}$ $\sim$ 0.8; section \ref{subsection: visibility}), again suggests that the major contribution to the Br$\gamma$ line in HD 100546 is not directly tracing accretion. Moreover, MA modelling of the Br$\gamma$ line considering stellar parameters similar to those of HD 100546 produces broader, double-peaked profiles that are not consistent with our observations \citep{Tambovtseva14}. We emphasize that this type of modelling assumes a disc truncation radius much larger than the one derived for HD 100546. This can strongly affect the broadening of the modelled line, as well as its double/single-peaked nature, depending on the spectral resolution assumed. Smaller magnetospheres could indeed be typical for HAeBe stars \citep{Cauley14}. Generally speaking, previous models that reproduce the Br$\gamma$ emission exclusively from magnetically channelled gas \citep{Muzerolle01,Kurosawa13} are not consistent with our observations.\\
In summary, whereas accretion in HD 100546 could be magnetospheric, the major contribution to the Br$\gamma$ emission comes from a Keplerian, probably flared, gaseous disc, and not from magnetically channelled gas. The accretion rate and the inner disc mass indicate that the observed gas should be replenished, perhaps from planet-induced flows from the outer to the inner disc as have been reported in similar objects. The peculiar characteristics of HD 100546 make it the perfect laboratory both to test planet formation and to understand how accretion proceeds in HAeBe stars.

\section*{Acknowledgments}
The authors thank the anonymous referee for his/her useful comments on the original manuscript, which helped us to improve the paper.\\
This research has made use of the  \texttt{amdlib-3.0.8} of the Jean-Marie Mariotti Center\footnote{Available at http://www.jmmc.fr/amberdrs}.\\
A.D.C. acknowledges support from CNPq (grant 308985/2009-5).\\
J.D.I gratefully acknowledges funding from the European Union FP7-2011 under grant agreement no. 284405.\\
R.G.V. acknowledges the support from FAPESP (grant 2012/20364-4).\\


\label{lastpage}
\end{document}